# Control Loop Feedback Mechanism for Generic Array Logic Chip Multiprocessor

[1]V.Karthikeyan, [2]V. J. Vijayalakshmi

[1]Department of Electronics and Communication Engineering,
SVS College Of Engineering, Coimbatore, India

[2]Department of Electrical and Communication Engineering,
Sri Krishna College of Engineering & Technology, Coimbatore, India

Email: [1]karthick77keyan@gmail.com, [2]vijik810@gmail.com

## ABSTRACT

Control Loop Feedback Mechanism for Generic Array Logic Chip Multiprocessor is presented. The approach is based on control-loop feedback mechanism to maximize the efficiency on exploiting available resources such as CPU time, operating frequency, etc. Each Processing Element (PE) in the architecture is equipped with a frequency scaling module responsible for tuning the frequency of processors at run-time according to the application requirements. We show that generic array logic Chip Multiprocessors with large inter-processor First In First Outputs (First In First Outs) buffers can inherently hide much of the Generic Array Logic performance penalty while executing applications that have been mapped with few communication loops. In fact, the penalty can be driven to zero with sufficiently large First In First Outs and the removal of multiple-loop communication links. We present an example mesh-connected Generic Array Logic chip multiprocessor and show it has a less than 1% performance (throughput) reduction on average compared to the corresponding synchronous system for many DSP workloads. Furthermore, adaptive clock and voltage scaling for each processor provides an approximately 40% power savings without any performance reduction.

## 1. INTRODUCTION

Dynamic Frequency Scaling (DFS) [5] is a widely used technique aimed at adjusting computational power to application needs. It is often associated with Dynamic Voltage Scaling (DVS) therefore enabling to achieve significant power reductions when computing demand is low; some cited benefits also comprise the reduction of thermal hot spots that participate in the accelerated aging of the circuits [4] due to the thermal stress. In core systems such as general purpose processors and high performance embedded processors, the operating system is responsible for dynamically adjusting the frequency of each processor to the current workload. This is facilitated by the presence of dedicated hardware monitors that the OS can rapidly access. In Linux based systems, two popular policies are used at kernel-level: on-demand and conservative. The on-demand governor switches to the highest available frequency whenever a load is detected whereas the conservative policy incrementally increases the frequency in a step-by-step fashion, yielding to better power savings at the expense of a lesser reactivity. Fast Fourier Transform (FFT) processor], the latency of a block of data will likely be lower with the source synchronous multi-word flow control method compared to the single transaction handshaking method due to its higher throughput. Dual-clock First in First Outs are well-suited to provide asynchronous boundary communication using the source synchronous multi-word flow control method. Single processor in the Generic Array Logic system Processors utilizes individual programmable ring oscillators that are configurable over a wide range of frequencies. Each processor also contains two dual-clocks First, In First Outs, which write and read in independent clock domains and reliably transfer data across the asynchronous boundaries. GAL's systems need to bring together the circuits amid clock domains to consistently [4] transport statistics. Clock phase edge alignment time for unmatched clocks and synchronization circuitry introduces a synchronization delay. This delay normally results in a reduction of performance (throughout). In this section, we discuss in depth principles behind how Generic Array Logic clocking affects system throughput and find several key architectural features which can hide the Generic Array Logic effects. Fullyavoiding any Generic Array Logic performance penalties is possible for the described Generic Array Logic chip multiprocessor. [2-3]

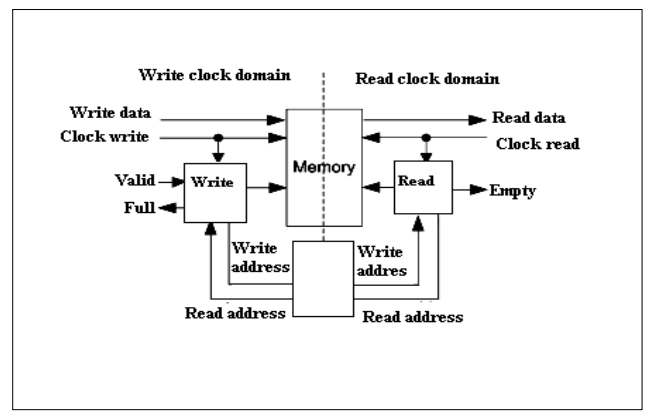

**Figure 1.** Clock domain

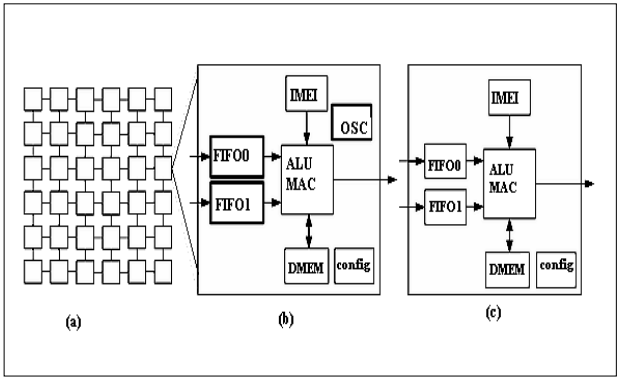

**Figure 2**. Reducing and Eliminating Performance Penalties of Generic Array Logic

## 2. PROPOSED WORK
### 2.1 Generic Array Logic

To simplify the discussion, in this section both Generic Array Logic and synchronous systems use the same clock frequencies. A. Related WorkSignificant previous research has studied the Generic Array Logic uniprocessor in which portions of each processor are located in separate clock domains. Results have shown Generic Array Logic Unipro-Caesars experience a non-negligible performance reduction compared to a corresponding synchronous uniprocessor. This value is sent to the frequency scaling module which will be responsible for scaling up and down the frequency of the processor to cope with application requirements. The procedure is then repeated and the obtained throughput gradually gets closer to the desired throughput. This is explained by the fact that after each iteration the error value is reduced assuming that the values of P, I and D have been correctly chosen. The system then calculates an error value which is obtained from the difference between the desired and obtained throughput. As output of the PID controller a frequency value is indicated.

## 3. IMPLEMENTATION OF MULTIPROCESSOR

It presents promising results regarding the adaptability of the system. We have demonstrated the efficiency of the proposed PID controller by presenting three different scenarios. For validating our approach we have implemented a multi-threaded version of the MJPEG decoder together with an ADPCM and FIR application which exchanges message using a message passing interface (MPI). The networking strategy between processors also strongly impacts the behavior of Generic Array Logic CMPs. Generic Array Logic chip multiprocessor with inter-processor and processor-memory communication through a shared global bus. This scheme provides very flexible communication, but places heavy demands on the global bus; thus, the system performance is highly dependent on the level and intensity of the bus traffic. Furthermore, this architecture lacks scalability since has increased global bus traffic will likely significantly reduce system performance under high traffic conditions or with a large number of processors.

## 4. RESULT ANALYSIS

The very small performance reduction of the Generic Array Logic chip multiprocessor motivates us to understand the factors that affect performance in GAL's style processors. The chain of events that allow synchronization circuit latency to finally affect application throughput. It is a complex relationship and several methods are available to hide the Generic Array Logic penalties. Increasing First In First Out Sizes: Increasing the First In First Out size will reduce First In First Out stalls as well as FIRST IN FIRST OUT stall loops, and hence increases system performance and reduce the Generic Array Logic performance penalty With a sufficiently large First In First Out, there will be no First In First Out full stalls and the number of First In First Out empty stalls can also be greatly reduced then the communication loop will be broken and no GAL's performance penalties will result. The top and middle subplots performance with different First In First Out sizes for the synchronous and Generic Array Logic systems, respectively. We assume all First In First Outs in the system are the same size and thus, their sizes are scaled together in this analysis.

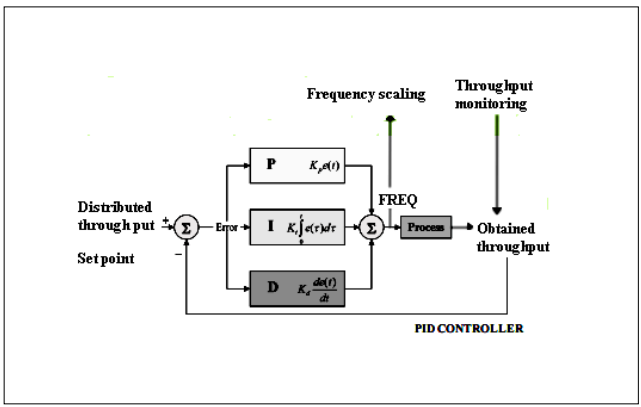

**Figure 3.** PID Controller

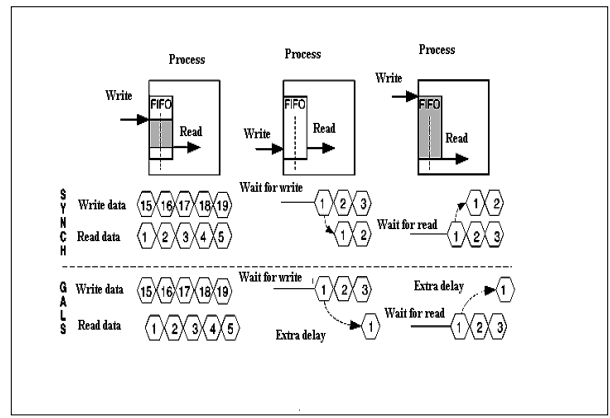

**Figure 4.** The result Synchronization circuit latency G6

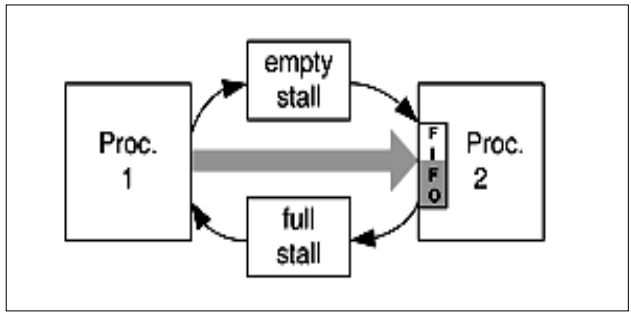

Figure 5. Eliminating Performance Penalties

## 5. CONCLUSION

In the projected work we solve the multithreading Dynamic memory location to spread out [5] to all other sequence Data presented in this paper are based on simulations of a fully-functional fabricated Generic Array Logic chip multiprocessor and physical designs based on the chip. Results from this work apply to systems with three key features as 1) Multi core processors (homogeneous and heterogeneous) operating in independent clock domains. 2) Source synchronous multi-word flow control for asynchronous boundary communication. 3) Distributed interconnect, such as a mesh. While the results certainly vary over different applications and specific architectures, systems with these features should still exhibit the following benefits over many workloads: good scalability, small performance reductions due to asynchronous communication overhead, and large potential power reductions from clock frequency and supply voltage scaling. Results show the system's capability of adapting to disturbing conditions. However, its use typically also introduces performance penalties due to the additional communication latency between clock domains

## 6. EXPERIMENTAL RESULTS

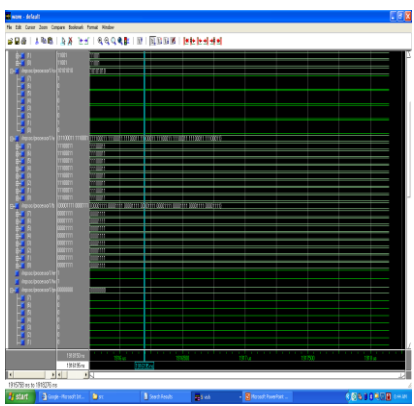

**Figure 6.** Multiprocessor Module

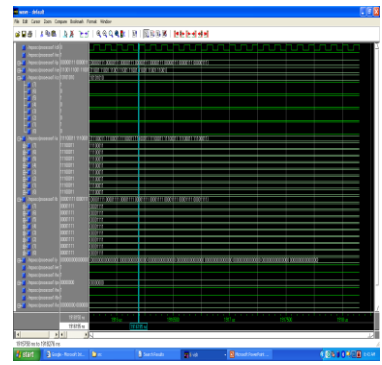

Figure 7. Generic Array Logic Module

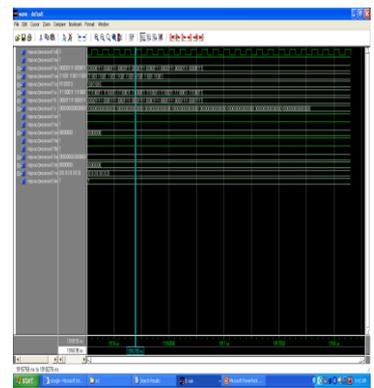

Figure 8. Multiprocessor with Generic Array Logic Module

## REFERENCES

[1] A. Acquaviva, A. Alimonda, S. Carta, and M. Pittau. Assessing task migration impact on embedded soft real-time streaming multimedia applications. EURASIP J. Embedded Syst., 2008:1–15, 2008.

[2] A. Alimonda, and S. Impact of task migration on streaming multimedia for embedded multiprocessors: A quantitative evaluation In ESTImedia, pages 59–64. IEEE, 2007.

[3] S. Bertozzi and Supporting task migration in multi-processor systems-on-chip: A feasibility study. In DATE '06. Proceed-ings, volume 1, pages 1–6, 2006

[4] S. Borkar, T. Kainik, S. Narendra, J. Tschanz, "Parameter variations and impact on circuits and microarchitec-ture," in Proc. IEEE Int. Conf. Des. Autom pp. 338–342, Jun. 2003.

[5] F. Clermidy. Dynamic and Distributed Frequency Assignment for Energy and Latency Constrained MP-SoC. On DATE'09, pages 1564–1567, Nice, France, 04 2009.

## AUTHOR PROFILES

**Prof V. Karthikeyan** has received his Bachelor's Degree in Electronics and Communication Engineering from PGP college of Engineering and Technology in 2003, Namakkal, India, He received a Masters Degree in Applied Electronics from KSR college of Technology and Erode in 2006 He is currently working as Assistant Professor in SVS College of Engineering and Technology, Coimbatore. He has about 8 years of Teaching Experience

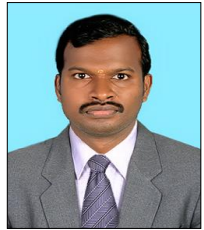

**Prof. V.J. Vijayalakshmi** has completed her Bachelor's Degree in Electrical & Electronics Engineering from Sri Ramakrishna Engineering College, Coimbatore, India. She finished her Masters Degree in Power Systems Engineering from Anna University of Technology, Coimbatore, She is currently working as Assistant Professor in Sri Krishna College of Engineering and Technology, Coimbatore She has about 5 years of teaching Experience.

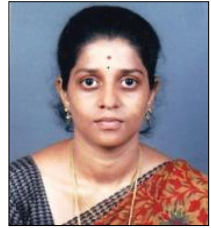